\documentclass[aip,jmp
,reprint 
,onecolumn,a4paper
,secnumarabic%
,amssymb, amsmath,nobibnotes, aps, showpacs, bm, verbatim]{revtex4-1}
\usepackage{docs,bm}%
\usepackage{tikz}
\usetikzlibrary{trees}
\colorlet{lightblue}{blue!20!}
\expandafter\ifx\csname package@font\endcsname\relax\else
 \expandafter\expandafter
 \expandafter\usepackage
 \expandafter\expandafter
 \expandafter{\csname package@font\endcsname}%
\fi

\usepackage[T1]{fontenc}

\def\Journal#1#2#3#4{{#4} {\it #1} {\bf #2}, #3 }
\newcommand{\w}[1]{\bm{#1}} 
\def\tho{\textrm{\TH}}
\def\thd{\tho '}
\def\et{\eth}
\def\etd{\eth '}
\def\ud{\textrm{d}}

\def\lc{\overline{\lambda}}
\def\muc{\overline{\mu}}
\def\tc{\overline{\tau}}
\def\pc{\overline{\pi}}
\def\pic{\overline{\pi}}
\def\kc{\overline{\kappa}}
\def\nc{\overline{\nu}}
\def\nuc{\overline{\nu}}
\def\rhoc{\overline{\rho}}

\def\sigmac{\overline{\sigma}}
\def\Sc{\overline{S}}
\def\tauc{\overline{\tau}}
\def\Pc{\overline{\Psi}}

\def\s{\sigma}
\def\restwo{(2)}
\def\resthree{(3)}
\def\resfour{(4)}

\newcommand{\smfrac}[2]{\textstyle{\frac{#1}{#2}}}

\newcommand{\be}{\begin{equation}}
\newcommand{\ee}{\end{equation}}
\newcommand{\ftn}{\cite}

\begin{document}

\title{Algebraically special Einstein-Maxwell fields}

\author{Norbert \surname{Van den Bergh}}
\email{norbert.vandenbergh@ugent.be}
\affiliation{Department of Mathematical Analysis FEA16, Gent University, Galglaan 2, 9000 Gent, Belgium}
\date{\today}
\begin{abstract}
The Geroch-Held-Penrose formalism is used to re-analyse algebraically special non-null Einstein-Maxwell fields, aligned as well as non-aligned, in 
the presence of a possible non-vanishing cosmological constant. 
A new invariant characterization is given of the 
Garc\'\i a-Pleba\'nski and Pleba\'nski-Hacyan metrics within the family of
aligned solutions and of the Griffiths metrics within the family of the non-aligned solutions. As a corollary also the double alignment of the 
Debever-McLenaghan `class $\mathcal{D}$' metrics with non-vanishing cosmological 
constant is shown to be equivalent with the shear-free and geodesic behavior of their Debever-Penrose vectors.\\

The final publication is available at Springer via http://dx.doi.org/10.1007/s10714-016-2171-x
\end{abstract}
\maketitle

\section{Introduction}

In the quest for exact solutions of the Einstein-Maxwell equations,
\begin{equation}\label{FEQS}
 R_{ab}-\frac{1}{2}R g_{ab}+\Lambda g_{ab} = F_{ac} {F_b}^c-\frac{1}{4}g_{ab}F_{cd}F^{cd},
\end{equation}
a large amount of research (see for example the reviews in \cite{GrifPod,Kramer} has been devoted to 
the study of so called \emph{aligned} 
Einstein-Maxwell fields, in which at least one of the principal null directions (PND's) of the electromagnetic field tensor $\w{F}$ is 
parallel to a PND of the Weyl tensor, also called a Debever-Penrose direction,
with main emphasis on the \emph{doubly aligned} Petrov type D solutions, in which both real PND's of $\w{F}$ 
are parallel to a corresponding double Weyl-PND. One of the prominent tools in these and related activities has been the Goldberg-Sachs 
theorem, which in its original form \cite{GoldbergSachs} says 
that a vacuum space-time is algebraically special if and only if it contains a shear-free geodesic null congruence (in an adapted Newman-Penrose tetrad, `$\Psi_0=\Psi_1=0$ if and only if $\kappa=\sigma=0$'). 
Goldberg and Sachs also proved that, if a space-time admits a complex null tetrad $(\w{k},\w{\ell},\w{m},\overline{\w{m}})$ such that 
$\w{k}$ is shear-free and geodesic 
and $R_{ab}k^a k^b = R_{ab}k^a m^b = R_{ab}m^a m^b = 0$ (as is the case when $\w{k}$ is a PND of the electromagnetic field tensor), then the Weyl tensor is algebraically special, with $\w{k}$ being a multiple
Weyl-PND.
While, for a null Maxwell field~\cite{Mariot,Robinson1961}, the Maxwell and Bianchi equations 
imply that  both conditions $\Psi_0=\Psi_1=0$ and $\kappa=\sigma=0$
are trivially satisfied (also in the presence of a possible cosmological constant), the situation is less straightforward in the non-null case. 
It still is true~\cite{GoldbergSachs,RobinsonSchild} that, when a PND $\w{k}$ of the electromagnetic field tensor is shear-free and 
geodesic, then the Weyl tensor is algebraically special, but the reverse no longer holds.
A key property in this respect is the Kundt-Tr\"umper theorem \cite{KundtTrumper}, which says that for an algebraically special aligned non-null Einstein-Maxwell space-time 
(with a possible non-0 cosmological constant and with $\w{k}$ the PND of $\w{F}$ coinciding with a
multiple Weyl-PND) one necessarily has 
\be \label{kappasigmaconds}
 \kappa(3 \Psi_2-2 |\Phi_1|^2)=0 \textrm{ and }
 \sigma(3 \Psi_2+2 |\Phi_1|^2)=0,
\ee
implying that the exceptional case $|\kappa|^2+|\sigma|^2 \neq 0$ can only occur for 
Petrov types II or D with
\begin{equation}\label{specPsi2}
 \frac{3}{2}\Psi_2=\pm |\Phi_1|^2.
\end{equation}
Subsequent research has been concentrated on the doubly aligned Petrov type D cases, in which the Kundt-Tr\"umper relations 
(\ref{kappasigmaconds}) also hold with $\kappa$ and $\sigma$ replaced by $\nu$ and $\lambda$. This 
culminated in the complete integration of the field equations for the Petrov type D doubly aligned non-null Einstein-Maxwell fields, with a 
possible non-0 cosmological constant:
\begin{itemize}
\item When $\frac{9}{4}\Psi_2^2- |\Phi_1|^4\neq 0$, both real PND's are geodesic and shear-free ($\kappa=\nu=\sigma=\lambda=0$) and, 
following Debever and McLenaghan\cite{DebeverMcLen1981}, I will refer to the corresponding set of solutions as the `class $\mathcal{D}$' space-times. They all admit 
at least a two-dimensional isometry group and count among their most famous members the Reissner-Nordstr\"om and Kerr-Newman metrics. 
Several authors \cite{Carter_PhysLett68,Carter_CMP68,Carter_BlackHoles1973,Debever1969,Debever1971,DebeverMcLen1981,DKM1983,DKM1984,GarciaSalazar1983,Garcia1984,Kinnersley_thesis,KowalczynskiPlebanski1977,PlebanskiDemianski1976,Plebanski1979} 
have independently contributed to the determination of class $\mathcal{D}$, beginning with Carter's seminal study of the separability of the Hamilton-Jacobi and Klein-Gordon equations and culminating in the
discovery \cite{PlebanskiDemianski1976} of Pleba\'nski and Demia\'nski's  
7-parameter metric for the non-null orbit solutions and Garc\'\i a's construction \cite{GarciaSalazar1983,Garcia1984} of a single metric form for 
both the non-null and null orbit solutions.
\item The case $\frac{3}{2}\Psi_2= |\Phi_1|^2$ was fully integrated in \cite{PlebHacyan79}. In the resulting `Pleba\'nski-Hacyan space-times'
$\w{k}$ and $\w{\ell}$ are respectively non-geodesic ($\kappa\neq0$) and geodesic 
($\nu =0$), while both are shear-free ($\sigma=\lambda=0$) and have vanishing complex divergence ($\rho=\mu=0$). The metric is given by
\be \label{PleHacmetric}
\ud s^2= 2\ud \zeta \ud \overline{\zeta}+2 \ud u\ud v+[\Lambda u^2+\zeta \overline{F}(v)+\overline{\zeta} F(v)] \ud v^2,
\ee
which, for an arbitrary function $F(v)$\footnote{Note1}
does not admit any isometries and only has an electromagnetic energy-momentum tensor of the correct sign
when $\Lambda<0$ .
\item The case $\frac{3}{2}\Psi_2= -|\Phi_1|^2$, which was overlooked in \cite{PlebHacyan79}, was dealt with in 
\cite{GarciaPleban}. Its unique solutions are
the `Garc\'\i a-Pleba\'nski metrics',\\
\be\label{GarciaPlebanmetric}
\ud s^2 = -\frac{1}{\Lambda}\left( e^{2z}{\omega^1}^2+e^{-2 z}{\omega^2}^2+\ud z^2-4 (\cosh z)^2{\omega^3}^2 \right),
\ee
with
\be
\omega^1+i \omega^2 = 2\frac{e^{iu}}{1-\zeta \overline{\zeta}}\ud \zeta \textrm{ and } \omega^3 = \ud u-i \frac{\zeta \ud \overline{\zeta}-\overline{\zeta}
\ud \zeta}{1-\zeta \overline{\zeta}}.
\ee
The corresponding space-times admit a 3 dimensional group of isometries; both $\w{k}$ and $\w{\ell}$ are geodesic ($\kappa=\nu=0$), shearing 
($\sigma\lambda \neq0$) and twisting, but non-expanding 
 ($\rho, \mu \in i \mathbb{R}$). Also this metric can only describe an Einstein-Maxwell space-time with an electromagnetic energy-momentum tensor of the correct sign when $\Lambda <0$.
\end{itemize}

In this paper I will have a closer look at the full set of algebraically special cases, both aligned (and hence by (\ref{specPsi2}) necessarily 
of Petrov type II) and non-aligned ones. First we will see that the 
Garc\'\i a-Pleba\'nski metrics are the unique members 
of the class of algebraically special and  aligned Einstein-Maxwell solutions for which the null direction $\w{k}$ is shearing. 
While doing so, I will correct an error in \cite{Kozarzewski} (cited also in \cite{Kramer} p.~409), claiming that, 
at least for $\Lambda=0$, the case $\kappa=0\neq \sigma$ admits no solutions.
This is true indeed but, as will become clear in section 2, the proof requires a much more subtle argumentation than the one presented in \cite{Kozarzewski}\footnote{Note2}
. The error occurs after relation (2.6)
when the case $\overline{\rho}=-\rho$ is dismissed by remarking that it ''leads to $\rho=0$''.
Most likely this conclusion was prematurely arrived at by inspection of 
the Newman-Penrose equation corresponding to our GHP equation (\ref{ghp1}): 
with $\kappa=\epsilon+\overline{\epsilon}=\Phi_0=0$ this equation reads $D\rho = \rho^2+\sigma\overline{\sigma}$, the real and imaginary parts of which only allow one to 
conclude that $\rho=\pm i |\sigma |$ and $D \rho =D \sigma = 0$.
In section 2 a correct proof of Kozarzewski's no-go claim will be provided, generalizing it to the case
$\Lambda \geqslant 0$ and showing that for $\Lambda<0$ the only allowed solutions are the doubly aligned Garc\'\i a-Pleba\'nski metrics with $\rho = \pm i |\sigma | \neq 0$.\\

In section 3 I will consider the algebraically special and  aligned Einstein-Maxwell fields for which the null direction $\w{k}$ is non-geodesic.
The general solution in this family so far is not known, but, remarkably, the Pleba\'nski-Hacyan metrics exhaust the sub-family characterised by the vanishing of 
the complex divergence of $\w{k}$.\\

Finally I prove in section 4 that an algebraically special Einstein-Maxwell solution possessing a shear-free and geodesic multiple Weyl-PND which is \emph{not} a PND of $\w{F}$ necessarily has vanishing cosmological 
constant and I give a characterization of the
sub-class of the Griffiths \cite{Griffiths} solutions. As a corollary of this theorem it follows that the `class $\mathcal{D}$'
metrics\cite{DebeverMcLen1981} are the unique Petrov type D Einstein-Maxwell solutions for which the real Weyl-PND's are both geodesic and shear-free and for which the cosmological constant is non-vanishing: in other words, the double alignment 
condition of the class $\mathcal{D}$ metrics with non-vanishing cosmological constant is a consequence of their multiple Weyl-PND's being geodesic and shear-free. Whether this property persists when $\Lambda=0$ is at present still 
an open problem, with only the Kundt case ($\rho=0$) so far having been dealt with.\\

In order to study these and related issues, linking kinematic properties of certain invariantly defined null directions (such as being geodesic and/or non-shearing) to algebraic properties
of the electromagnetic field tensor or of the Weyl tensor, it is natural to use the Geroch-Held-Penrose (GHP) formalism \cite{GHP}. In this formalism only a pair of null \emph{directions} is singled out at each point 
rather than an entire null tetrad, as is the case in the Newman-Penrose \cite{NP} formalism. The resulting formalism is covariant with respect to rotations of the spatial basis vectors and boosts of the real null
directions and, as such, is `halfway' between a fully covariant approach and the NP spin-coefficient approach and leads to considerably simpler equations with 
fewer complex variables.

Throughout I will use the sign conventions and notations of {Kramer} \S 7.4, with the tetrad basis vectors taken as $\w{k},\w{\ell},\w{m},
\overline{\w{m}}$ with $- k^a \ell_a = 1 = m^a
\overline{m}_a$. However, in order to ease comparison with the (more familiar) Newman-Penrose formalism, I will write primed variables, such as $\kappa',\sigma',\rho'$ and $\tau'$, as their 
NP equivalents $-\nu,-\lambda, -\mu$ and $-\pi$.  For completeness the resulting weights, commutators, GHP, Bianchi and Maxwell equations are 
presented in an Appendix.

Finally note that, at least when the electromagnetic field is non-null, the pair of null directions $\w{k},\w{\ell}$ can
always be invariantly defined, aligning for example $\w{k}$ with a PND of $\w{F}$ and null-rotating about $\w{k}$ such that $\Phi_1$ is the only non-vanishing 
component of the Maxwell spinor. Obviously this choice is not unique (for example in the non-aligned case it will be preferable to align $\w{k}$
with the multiple Weyl-PND and to null-rotate about $\w{k}$ such that $\Phi_1=0$ and $\Phi_0 \Phi_2\neq 0$), but it is important to realise
that any ensuing well-weighted GHP relations, such as $\kappa=0, \rho =0, \ldots$ are automatically \emph{geometrically invariant statements}. 

\section{Aligned electrovacs with a shearing multiple Debever-Penrose vector}
Let us first, for the sake of completeness, re-confirm the well-known fact~\cite{GoldbergSachs,RobinsonSchild} that if a PND $\w{k}$ of the electromagnetic field tensor $\w{F}$ is shear-free and 
geodesic, then the Weyl tensor is algebraically special with $\w{k}$ being the multiple Weyl-PND: with $\w{k}$ a PND of $\w{F}$ ($\Phi_0=0$) satisfying $\kappa=\sigma=0$,
it follows from (\ref{ghp2}) that $\Psi_0=0$ and hence $\w{k}$ is also a Weyl-PND. When
$\w{F}$ is non-null 
we can null-rotate about $\w{k}$ such that also $\Phi_2=0$. 
The integrability conditions for the Maxwell equations (\ref{max1}, \ref{max2}) 
simplify then with (\ref{ghp3}, \ref{ghp8}) to $\Phi_1 \Psi_1=0$. It follows that $\Psi_1=0$ and hence $\w{k}$ is a multiple Weyl-PND.\\

Now let us consider the reverse situation and assume that the Weyl tensor is algebraically special, with the multiple Weyl-PND $\w{k}$ being also a PND of the non-null electromagnetic field tensor: by means of 
a suitable null rotation about $\w{k}$ the null tetrad can be chosen
such that $\Psi_0=\Psi_1=\Phi_0=\Phi_2=0$. The Maxwell equations (\ref{max1}, \ref{max2}') 
and (\ref{max2}) become
\begin{equation}
 \tho \Phi_1=2 \rho \Phi_1, \ \et \Phi_1 =2\tau  \Phi_1,\ \etd \Phi_1 =- 2\pi  \Phi_1,
\end{equation}
while the Bianchi equations (\ref{bi1}, \ref{bi2}) 
yield the Kundt-Tr\"umper relations (\ref{kappasigmaconds}).\\
In the present paragraph we consider the case where $\w{k}$ is shearing and hence
\begin{equation}
\kappa=0,\ \Psi_2 = -\frac{2}{3} |\Phi_1|^2 .
\end{equation}
Bianchi equation (\ref{bi3}) simplifies now to $|\Phi_1|^2 (\rho+\rhoc)=0$, implying that $\w{k}$ is non-expanding. From the real part of GHP equation (\ref{ghp1}) it 
follows that $\rho$ is related to $\sigma$ by  
\be\label{rhosigmarel}
 \rho = \pm i |\sigma|
\ee
and hence, by (\ref{ghp1}, \ref{ghp2}, \ref{ghp3}) 
\be \label{Drhosigmatau}
\tho \rho=0,\ \tho \sigma =0,\ \tho \tau = \rho(\tau+\pc)+\sigma (\tc+\pi).
\ee 
From (\ref{bi4},\ref{bi4}') one finds then
\be
\Psi_3 = -\frac{2}{3} \frac{|\Phi_1|^2}{\sigma} (2\tau+\pc),
\ee
and
\be
\tho \pi = 2(\pc \sigmac -\tc \rho) .
\ee
Acting with the $[\et,\, \tho]$ commutator on $\Phi_1$ now leads to an expression for $\et \rho$, which, 
together with (\ref{ghp8}) and the $\et$ derivative of (\ref{rhosigmarel}), yields
\begin{eqnarray}
\et \rho &=& \sigma (\tc+2 \pi)+2\rho \tau,\label{et_rho}\\
\etd \sigma &=& \sigma (\tc+2\pi),\label{etdsig}\\
\et \sigma &=& \sigma(3\tau-2\pc)-2\rho \frac{\sigma}{\sigmac}(\tc+2 \pi).\label{etsig}
\end{eqnarray}
The remaining integrability conditions for the Maxwell equations can then be written as
\begin{eqnarray}
 \thd \tau+\eth \mu &=& \rho \nc+\lc \pi,\label{maxint3}\\
 \thd \pi -\etd \mu &=& \lambda \tau -\nu \rho-\mu\tc-\muc\pi,\label{maxint4}\\
 \eth \pi+\etd \tau &=& \rho (\mu+\muc),\label{maxint5}\\
 \tho \mu+\thd \rho &=& \pi\pic-\tau\tc.\label{maxint6}
\end{eqnarray}

Next we apply the $[\et,\, \tho]$ commutator to $\sigma$, to obtain an expression for $\tau$,
\be
\tau= \smfrac{1}{3}\pic+\frac{4}{3}\frac{\rho\pi}{\sigmac},
\ee
which together with its $\tho$ derivative leads to
\be
\sigma \pi -\rho \pc =0 \label{pic_expr}.
\ee
This suggests the definition of an auxiliary variable $S$ (by (\ref{rhosigmarel}) one has $|S|=1$, while $\textrm{weight}(S)=[2,-2]$) such that $\sigma= S \rho$ and enabling us to conclude from (\ref{pic_expr}) that
\be
\tau=-\pc,\  \pc= S \pi .
\ee
By (\ref{Drhosigmatau},\ref{et_rho}--\ref{etsig}) one has then
\begin{equation}
\tho S =0,\ \eth S=-S (3 \pic -S \pi),\ \etd S=3 S \pi-\pic ,
\end{equation}
after which (\ref{bi2}',\ref{bi3},\ref{bi3}') yield
\begin{eqnarray}
\eth \pi &=& S \pi^2 +2 \rho (\mu+\muc)-\frac{3}{2} S \rho^2 \Psi_4 |\Phi_1|^{-2}, \label{d1piexpr}\\ 
\etd \pi &=& -\pi^2-2 S^{-1} \rho (\mu+\muc)-\frac{3}{2} S^{-2} \rho^2 \overline{\Psi_4} |\Phi_1|^{-2},\\
\tho \Psi_4 &=& \rho (\Psi_4 -S^{-2}\overline{\Psi_4})-\frac{4}{3} S^{-1}(\mu+\muc)|\Phi_1|^2.
\end{eqnarray}
Herewith (\ref{maxint5}) reduces to
\be\label{max5int}
2 (\mu+\muc){|\Phi_1|^2}+\rho(S^{-1}\overline{\Psi_4}-S\Psi_4)= 0,
\ee
while expressing that $\thd (S\Sc)=0$ (with $\thd S$ evaluated from (\ref{ghp4d})) and simplifying the result by means of (\ref{max5int}), one 
finds that
\be
\mu+\muc = 0
\ee
(i.e.~also $\w{\ell}$ is non-expanding) and hence $\overline{\Psi_4}  = S^2 \Psi_4$.\\
Furthermore, calculating $\thd \rho$ from (\ref{ghp5d}) and expressing that $\thd(\rho+\rhoc)=0$ gives us an equation from which we can obtain
$\Psi_4, \thd \rho$ and $\thd S$:
\begin{eqnarray} 
\Psi_4 &=& \frac{|\Phi_1|^2}{18\rho^2} \left[ 24\pi^2+6\rho (\lambda -S^{-2}\lc)-S^{-1} (8 |\Phi_1|^2+12 \rho \mu -R) \right], \label{Psi4expr} \\
\thd \rho &=& -\frac{\rho}{2}(S\lambda +S^{-1}\lc ),\\
\thd S &=& S \left[ -2 \mu +S \lambda -S^{-1}\lc+\frac{1}{\rho}(2 S \pi^2-\smfrac{2}{3}|\Phi_1|^2+\smfrac{1}{12} R) \right].
\end{eqnarray}
Next we solve (\ref{bi1}', \ref{bi2}') together with (\ref{maxint6}) and the  $[\thd,\, \tho]\rho,\ [\eth,\, \thd ] S$ commutator relations for $\thd \pi$, $\eth \mu$, $\tho \mu$, $\tho \lambda$,
$\eth \lambda$ and $\etd \lambda$:
\begin{eqnarray}
  \thd \pi &=& -\smfrac{3}{2} S \lambda \pi+\smfrac{1}{2}S^{-1}\rho \nc+S^{-1}\lc\pi+\smfrac{1}{2}\mu\pi-\smfrac{5}{2}\nu\rho +( \smfrac{4}{3}|\Phi_1|^2 
-3 S \pi^2 -\smfrac{1}{8} R)\frac{\pi}{\rho},\label{d3piexpr}\\
  \eth \mu &=& -\smfrac{5}{2} S\rho\nu-\smfrac{1}{2}\lambda\pi S^2-\smfrac{3}{2}S \mu\pi+\smfrac{3}{2}\rho\nc+\lc \pi+(\smfrac{2}{3}  |\Phi_1|^2-S\pi^2-\smfrac{1}{24} R)\frac{S \pi}{\rho},\\
  \tho \mu &=& \smfrac{1}{2}\rho (S\lambda+S^{-1} \lc),\\
  \tho \lambda &=& \smfrac{1}{2}\rho (\lambda +S^{-1}\lc)-2 \pi^2-\smfrac{1}{12} S^{-1}(R-8 |\Phi_1|^2), \\
  \eth \lambda &=& -\smfrac{5}{2}(\mu\pi+\nu\rho)+\smfrac{3}{2}S \lambda\pi+\smfrac{7}{2} S^{-1}\rho\nc +(S \pi^2 -\smfrac{8}{3}|\Phi_1|^2+
\smfrac{1}{24}R)\frac{\pi}{\rho},\label{d1lambdaexpr}\\
  \etd \lambda &=& \smfrac{1}{2}S^{-1}(\mu\pi+\rho \nu)+\smfrac{3}{2}S^{-2}(3\rho\nc- \lc \pi )-3\lambda\pi +S^{-1}(3 S\pi^2-\smfrac{8}{3} |\Phi_1|^2+\smfrac{1}{8} R)\frac{\pi}{\rho},
\end{eqnarray}
after which an expression for the spin coefficient $\mu$ follows from the $[\etd,\, \eth ]S$ commutator relation,
\be\label{mu_expr}
\mu = \frac{1}{2}(S\lambda-S^{-1} \lc)+\frac{1}{36\rho}(24 S \pi^2-32 |\Phi_1|^2+R).
\ee
Of the Maxwell integrability conditions there only remains now the $[\etd,\, \thd]\Phi_1$ relation, namely
\be\label{eq1}
6 (S \lambda +S^{-1} \lc)\rho \pi +\pi (8 |\Phi_1|^2-24 S \pi^2-R) -36\rho^2(\nu- S^{-1}\nuc) = 0,
\ee
which with GHP equation (\ref{ghp8}') gets simplified to the key algebraic equation
\be\label{pi_key}
\pi (8|\Phi_1|^2-24 S \pi^2-R)=0.
\ee
Herewith all Bianchi equations and GHP equations (except those involving the derivatives of $\nu$) are identically satisfied. At this stage 
the Weyl spinor components are given by, using (\ref{mu_expr}) to simplify (\ref{Psi4expr}),
\begin{eqnarray}
 \Psi_0 &=& \Psi_1=0,\\
 \Psi_2 &=&  -\smfrac{2}{3}|\Phi_1|^2,\\
 \Psi_3 &=& \smfrac{2}{3}\frac{\pi}{\rho}|\Phi_1|^2 , \\
 \Psi_4 &=& \frac{|\Phi_1|^2}{27  \rho^2}S^{-1}(24 S \pi^2+4|\Phi_1|^2+R). \label{II_Psi4}
\end{eqnarray}
I now show that the case $\pi\neq 0$ is inconsistent, while the case $\pi=0$ leads to the Garc\'\i a-Pleba\'nski metrics.
\subsection{$\pi\neq 0$}
When $\pi\neq 0$ we can use (\ref{pi_key}) to rewrite (\ref{eq1}) as
\be
\pi(S\lambda+S^{-1}\lc)-6\rho (\nu-S^{-1} \nc)=0,
\ee
while (\ref{mu_expr}) becomes
\be \label{forgmu}
\mu=\smfrac{1}{2}(S\lambda-S^{-1}\lc)-\smfrac{2}{3}|\Phi_1|^2\rho^{-1}.
\ee
Taking the $\eth$ derivative of (\ref{forgmu}) and eliminating $\pi$ from the resulting equation and (\ref{pi_key}), we also find
\be\label{mu_rel}
3\rho(S\lambda-\Sc\lc)-4|\Phi_1|^2=0
\ee
and hence, by (\ref{forgmu}), $\mu=0$. The $\tho$ derivative of (\ref{mu_rel}) yields then $S\lambda+S^{-1} \lc=0$ and hence, by (\ref{mu_rel}), $\lambda=
\smfrac{2}{3}|\Phi_1|^2 S^{-1} \rho^{-1}$. Substituting this in the expression (\ref{d1lambdaexpr}) for $\eth \lambda$ one obtains 
$\nu=\smfrac{2}{3}|\Phi_1|^2\pi \rho^{-2}$, which together with GHP equation (\ref{ghp2}') leads to the contradiction $|\Phi_1|^2\pi=0$.

\subsection{$\pi=0$}
Substituting $\pi=0$ in (\ref{d1piexpr},\ref{II_Psi4}) one obtains 
\be
R \, (=4 \Lambda) =-4|\Phi_1|^2 \label{negativeLambda}
\ee
and 
\be \nc-5 S \nu=0, 
\ee 
whence $\nu=0$. While (\ref{negativeLambda}) proves Kozarzewski's no-go claim for $\Lambda=0$, it also generalizes it to the case $\Lambda \geqslant 0$. Furthermore, when $\Lambda < 0$ 
the only non-0 Weyl spinor component is $\Psi_2=R/6$ and we are in the doubly aligned 
situation with both Weyl-PND's $\w{k}$ and $\w{\ell}$ being geodesic and non-expanding. 
The field equations and Maxwell equations have been completely integrated in this case by Garc\'\i a and 
Pleba\'nski~\cite{GarciaPleban} and the resulting metric is given by (\ref{GarciaPlebanmetric}), admitting a 3D isometry group.

\section{Aligned electrovacs with a non-geodesic multiple Debever-Penrose vector}
We again consider the case where the Weyl tensor is algebraically special, with the multiple Weyl-PND $\w{k}$ being also a 
PND of the non-null electromagnetic field tensor, but now we take $\kappa \neq0$ and hence, by (\ref{kappasigmaconds}),
\begin{equation}
\sigma=0 \textrm{ and } \Psi_2 = \frac{2}{3} |\Phi_1|^2 .
\end{equation}

Bianchi equation (\ref{bi4}) now reduces to $|\Phi_1|^2 (\tau-\pc)=0$, whence $\tau=\pc$, after which (\ref{bi3}) implies
\be
\Psi_3=\frac{2}{3}\frac{2\rho-\rhoc}{\kappa}|\Phi_1|^2, \label{sec2_psi3}
\ee
showing that solutions cannot be of Petrov type III or N when $\rho\neq0$.

GHP equations (\ref{ghp2},\ref{ghp8},\ref{ghp4d}) and Bianchi equation (\ref{bi3}) yield then 
\begin{eqnarray}
 \eth \kappa &=& 0 ,\\
 \eth \rho &=& \pc (\rho-\rhoc)+\kappa (\mu-\muc),\\
 \etd \rho &=& -2 \pi \rhoc -2 \kc \muc,\label{s2_d2rho}\\
 \etd \pi &=& \pi^2+\lambda \rhoc -\nu \kc.
\end{eqnarray}
The integrability conditions for the Maxwell equations furthermore give an expression for $\tho \pi$,
\be
\tho \pi = \kc(2\muc -\mu)+2\pi \rhoc ,
\ee
together with four extra relations
\begin{eqnarray}
 \eth \mu+\thd \pc &=& \nuc \rho+\lc \pi,\\
 \etd \mu -\thd \pi &=& \pi(\muc +\mu)-\lambda \pic+\nu \rho,\\
 \eth \pi +\etd \pic &=& \rho \muc -\rhoc \mu, \\
 \tho \mu +\thd \rho &=& 0,
\end{eqnarray}
with which GHP equations (\ref{ghp3},\ref{ghp4d}') imply
\begin{eqnarray}
 \thd \kappa &=& \kappa (2\mu-\muc),\\
 \tho \lambda &=& \lambda(\rho+\rhoc)+2\pi^2-2\nu\kc .
\end{eqnarray}

So far it has not been possible to complete the analysis of this case. However, it is easy to see that the Pleba\'nski-Hacyan metrics completely
exhaust the non-diverging subfamily of solutions ($\rho=0$) and have  $\Lambda <0$, whereas for
$\Lambda \geqslant 0$ no solutions exist. In fact, when $\rho=0$ (\ref{s2_d2rho}) and $\kappa\neq0$ imply 
that also the second PND $\w{\ell}$ of the electromagnetic field is 
non-diverging ($\mu=0$): herewith (\ref{sec2_psi3}) and (\ref{bi4}') show that $\Psi_3=\Psi_4=0$, so that we are in the doubly aligned Petrov type D case. 
This suffices already to conclude that the only possible solutions are the
Pleba\'nski-Hacyan metrics, as this case was completely integrated in \cite{PlebHacyan79}, modulo the error corrected in \cite{GarciaPleban}. 
However, it is instructive to provide a short and coordinate independent proof of the essential step in 
\cite{PlebHacyan79}, namely the part in which the authors prove that $\Gamma_{423}=0$, leading to the conclusion that
$\w{\ell}$ not only is shear-free ($\lambda=0$ being an immediate consequence of (\ref{bi2}')), but also is geodesic ($\nu=0$):\\

First observe that the $[\eth,\, \thd]$ commutator applied to $\kappa$ implies $\nc \tho  \kappa=0$. Assuming $\nu \neq 0$ leads then to an inconsistency, as we would have 
\be
\tho \kappa=0, \ \etd \kappa= \kc\pc-\kappa\pi,
\ee
with the second relation being obtained from (\ref{ghp1}).
Herewith the $[\tho,\, \thd]\kappa$ commutator relation yields
\be
\kappa (\kc \nuc +3\kappa\nu-6\pi\pic+\smfrac{1}{6}R-\smfrac{20}{3}|\Phi_1|^2)+2\kc \pc^2=0,
\ee
which, with $\nu \kc =\pi^2$ (obtained from (\ref{ghp4d}')) simplifies to 
\be\label{R_eq1}
|\kappa|^2(40|\Phi_1|^2-R)-18(\kappa \pi-\kc\pic)^2=0.
\ee
On the other hand (\ref{ghp5d}'-$\overline{\ref{ghp5d}}$) reads
\be\label{R_eq2}
 |\kappa|^2(8|\Phi_1|^2+R)-6(\kappa \pi-\kc\pic)^2=0,
\ee
which, when added to (\ref{R_eq1}), results in 
\be
2|\kappa\Phi_1|^2 - (\kappa\pi-\kc\pc)^2 =0.
\ee
The left hand side being positive definite, this shows that the case $\nu\neq 0$ is inconsistent. We therefore have $\nu=\pi=0$ and hence, by GHP equation (\ref{ghp5d}), 
$R = 4\Lambda =-8|\Phi_1|^2 < 0$. 

\section{Electrovacs with a shear-free and geodesic multiple Debever-Penrose vector}
We now consider algebraically special (non-conformally flat) Einstein-Maxwell fields with a possible non-zero cosmological constant for which the 
multiple Weyl-PND $\w{k}$ is geodesic and shear-free 
($\Psi_0=\Psi_1=\kappa=\sigma=0$) 
and for which $\w{k}$ is not parallel to a PND of $\w{F}$ ($\Phi_0\neq0$). Choosing a null-rotation about $\w{k}$ such that $\Phi_1=0$, it 
follows that $\Phi_2\neq 0$ \footnote{with $\Phi_2 = 0$ $\w{\ell}$ would be geodesic and shear-free;
according to the generalised Goldberg-Sachs theorem we would have then $\Psi_3=\Psi_4=0$ and the Petrov type would be D, in which case 
\cite{DebeverVdBLeroy,VdB1989} the only null Einstein-Maxwell solutions are given by the (doubly aligned)
Robinson-Trautman metrics}.

The Maxwell equations (\ref{max1},\ref{max2}) and Bianchi equations (\ref{bi1}-\ref{bi4}) yield then
\begin{eqnarray}
\eth \Phi_0  &=& 0,\\
\etd \Phi_0  &=& -\pi \Phi_0,\\
\tho \Phi_0  &=& 0,\\
\thd \Phi_0  &=& -\mu \Phi_0,\\
\eth \Phi_2   &=&  -\nu \Phi_0+\tau \Phi_2,\\
\tho \Phi_2  &=& -\lambda\Phi_0+\rho \Phi_2,\\
\eth \Psi_2  &=& -\pi \Phi_0\overline{\Phi_2}+3\tau \Psi_2,\\
\tho \Psi_2   &=&  \mu |\Phi_0|^2+3 \rho \Psi_2,
\end{eqnarray}

after which the commutators $[\etd,\, \eth], [\etd,\, \tho], [\eth, \, \thd]$ and $[\thd,\, \tho]$ applied to $\Phi_0$ give

\begin{eqnarray}
 \eth \pi   &=&  (3\rho -\rhoc) \mu -2\Psi_2+\frac{R}{12},\\
 \tho \pi   &=&  3\rho \pi,\\ \label{dpi}
 \eth \mu   &=&  \lc \pi +3 \mu \tau,\\
 \tho \mu   &=&  \pi(\pc+3 \tau) +2 \Psi_2-\frac{R}{12}. \label{dmu}
\end{eqnarray}

Herewith GHP equation (\ref{ghp5d}') becomes a simple algebraic equation for $\Psi_2$,
\be
\Psi_2 = \rho \mu-\tau\pi+\frac{R}{12},
\ee
the $\tho$ derivative of which (using (\ref{dpi},\ref{dmu},\ref{ghp1},\ref{ghp3})) results in $\rho R=0$.\\

As $\rho =0$ would imply $\Phi_0=0$, this shows that an algebraically special Einstein-Maxwell solution possessing a shear-free and 
geodesic multiple Weyl-PND which is \emph{not} a PND of $\w{F}$ necessarily has a vanishing cosmological 
constant\footnote{Note3}. The corresponding class of solutions is non-empty: imposing the additional restriction that $\pi=0$ one can deduce that $\rhoc=\rho$ and 
$\muc=\mu$, together with a $[-2,0]$--weighted relation $\Psi_3=\rho \nu+\mu \tc-\tau \lambda$.
It is then straightforward to construct a Newman-Penrose null-tetrad with the additional restrictions $\alpha=\beta=\epsilon=0$ and $\mu=2\gamma$: the unique solutions in this case are the Griffiths \cite{Griffiths} metrics, containing as 
special cases the metrics \cite{CahenLeroy,CahenSpelkens,Griffiths1983,SzekeresJ1966}. At first sight one would expect the class $\pi\neq 0$ to 
admit a larger set of solutions, but this is by no means guaranteed (compare with section 2.1): anyway no explicit examples appear to be known. The only property which is easy to demonstrate --though 
somewhat tedious to be included in the present paragraph-- is that no solutions exist for which $\w{k}$ is non-expanding ($\Re(\rho)\neq0$).\\

As a corollary of the above result we also obtain a new characterization of the class $\mathcal{D}$ metrics\cite{DebeverMcLen1981} with non-vanishing cosmological constant: if one of the multiple 
Weyl-PND's a Petrov type D Einstein-Maxwell solution would 
be non-aligned with a PND of $\w{F}$ then $R=0$. In other words, for a non-vanishing cosmological constant, the double alignment property of the 
class $\mathcal{D}$ metrics is a consequence of their Weyl-PND's being geodesic and shear-free.

It is tempting to conjecture that this same conclusion also will hold when $\Lambda=0$. However it has not been possible so far to prove this, except for the special case of the Kundt space-times (i.e.~in which the 
Weyl-PND $\w{k}$ has vanishing complex divergence). As the proof for this particular case is again quite tedious and little illuminating, I prefer to postpone this part to
a possible later and more general publication.

\section{Discussion}
Most results dealing with exact solutions for (algebraically special) Einstein-Maxwell fields have been obtained in the past introducing 
special coordinate systems, usually adapted to geodesic and/or shear-free null-congruences, or by 
imposing, sometimes haphasard looking, restrictions on
the spin coefficients of a Newman-Penrose tetrad. This not only has turned their comparison and classification into an awkward procedure, 
often involving sophisticated computer algebra packages, but also has made it difficult for researchers and students entering the field to recognize
the blanks which remain to be filled in.
As furthermore the ``Exact Solutions book''\cite{Kramer} touches the subject of non-aligned Einstein-Maxwell fields, or of aligned fields with 
a non-geodesic or shearing multiple PND, 
only superficially,
I present below the known results, together with the ones obtained in the previous sections, schematically.
From these diagrams I exclude the conformally flat case, as it implies\cite{PodolskyOrtaggio} 
$\Lambda=0$ with 
the only non-null member being therefore\cite{Kramer} the Bertotti-Robinson\cite{LeviCivita,Bertotti,Robinson1959} metric and the null members being given by 
a special class of plane waves\cite{BaldwinJeffery,McLenTariqTupper}.

In the figures below labels (1), (2), \ldots next to vertical arrows refer to results obtained in the corresponding previous sections, 
$\nexists$ indicates that no solutions are allowed, while a question mark indicates that no solutions are known to occur in the literature. Roman capitals 
II, III, N, D refer to the Petrov types and $`1\times`$ or $`2\times`=$ indicate singly aligned or doubly aligned solutions.

\bigskip 

\begin{figure}[!htb]
\small{
\begin{center}
\begin{tikzpicture}[level distance=4em,
  every node/.style = {shape=rectangle, rounded corners,
    draw, align=center}]
    [\tikzstyle{level 1}=[sibling distance=20em],\tikzstyle{level 2}=[sibling distance=8em],\tikzstyle{level 3}=[sibling distance=6em],\tikzstyle{level 4}=[sibling distance=2.5em]]
    \node (a) {non-O, non-null, aligned}
    child [sibling distance=22em] { node (b) {$|\kappa|^2+|\sigma|^2\neq 0$}
      child [sibling distance=10em] {node (b1) {$\sigma=0\neq \kappa$}
        child { node (b11) {$\rho=0$}
          child { node (b111) {$\Lambda<0$}
            child { node (b211) [label={[draw=none]below: {\begin{tabular}{c} type D, $2\times$ \end{tabular}}}] {PH\cite{PlebHacyan79}}
            edge from parent [->] node [left,draw=none] {\tiny \resthree }
            }
          edge from parent [->] node [left,draw=none] {\tiny \resthree }
          }
        }
        child { node (b12) {$\rho \neq 0$}
          child { node (b121) {II}
            child { node (b1211) {?}
            edge from parent [->] node [right,draw=none] {}
            }
          }
          child { node (b122) {III}
            child { node (b1211) {$\nexists$}
            edge from parent [->] node [left,draw=none] {\tiny \resthree }
            }
          }
          child { node (b121) {N}
            child { node (b1211) {$\nexists$}
            edge from parent [->] node [left,draw=none] {\tiny \resthree }
            }
          }
        }
      }
      child { node (b2) {$\kappa=0\neq \sigma $}
        child { node (b21) {$\Lambda < 0, \rho = \pm i \sigma$}
          child { node (b211) [label={[draw=none]below: {\begin{tabular}{c} type D, $2\times$ \\ $\nu=0\neq \lambda$  \end{tabular}}}] {GP\cite{GarciaPleban}}
          edge from parent [->] node [left,draw=none] {\tiny \restwo}
          }
        edge from parent [->] node [left,draw=none] {\tiny \restwo}
        }  
      }
      child { node (b3) {$\kappa\sigma \neq 0$}
        child { node (b31) {$\nexists$}
        edge from parent [->] node [right,draw=none] {\tiny Ref.~\cite{KundtTrumper}} 
        }  
      }
    }
    child [sibling distance=20em] { node (c) {$\kappa=\sigma=0$}
      child [sibling distance=9em] { node (c1) {$2\times$}
        child [sibling distance=5em] { node (c11) {II}
          child { node (c111) [label={[draw=none]below: {\tiny $\Lambda \neq 0$?}}] {L\cite{Leroy79}}
          edge from parent [->] node [right,draw=none] {}
          }
        }
        child [sibling distance=5em] {node (c12) {D}
          child { node (c121) [label={[draw=none]below: {}}] {$\mathcal{D}$\cite{PlebanskiDemianski1976,Garcia1984}, PH'}
          edge from parent [->] node [right,draw=none] {\tiny Ref.~\cite{DebeverMcLen1981}}
          }
        }
        child [sibling distance=5em] {node (c13) {III}
          child { node (c131) {$\nexists$}
          edge from parent [->] node [right,draw=none] {\tiny Ref.~\cite{Kramer}}
          }
        }
      }
      child {node (c1) [label={[draw=none]below: {\tiny Ref.~\cite{Kramer}}}] {$1\times$}
      }
    }   
    ;
    \end{tikzpicture}
\end{center} 
}
\caption{\label{fig:aligned}}{Algebraically special non-nul Einstein-Maxwell solutions for which the Weyl-PND $\w{k}$ is a PND of $\w{F}$, the null-rotation about $\w{k}$ 
being chosen such that $\Phi_1$ is the only non-vanishing 
component of the Maxwell spinor; L $=$ Leroy, PH $=$ Pleba\`nski-Hacyan, GP $=$ Garc\`\i a-Pleba\`nski.}
\end{figure}

In Fig.~1 the sub-tree corresponding to the singly aligned Einstein-Maxwell
solutions with a shear-free and geodesic multiple Weyl-PND (as well as the null fields, which automatically obey this condition) is not included, as little or no progress has been made in this area since the early 
nineties\cite{Kramer,Nurowski}, with the bulk of the material contained in \cite{Kramer}. The double occurence of the Pleba\`nski-Hacyan metric in Fig.~1
(once as PH and once as PH', the GHP-primed version of PH with the roles of $\w{k}$ and $\w{\ell}$ interchanged) is due to the fact that PH is 
doubly aligned with only one null-ray being geodesic and both being shear-free.
The absence of doubly aligned Petrov type III solutions in Fig.~1 has been noticed already in \cite{Kramer} for $\Lambda=0$, but can 
easily be seen to hold also for $\Lambda\neq 0$. \cite{Kramer} also mentions that the Leroy metric\cite{Leroy79} is the unique Einstein-Maxwell 
solution of Petrov type II with $\Lambda=0$, in which both the Weyl PND's $\w{k}$ and $\w{\ell}$ are also PND's of $\w{F}$ and in which the multiple PND $\w{k}$ is geodesic and shear-free; this is not clear at all, as 
Leroy's solutions were only obtained in the non-radiative sub-case. It furthermore remains to be checked whether 
this still holds --for a suitable generalisation of the Leroy 
metric-- when $\Lambda \neq 0$.

\begin{figure}[!htb]
\small{
\begin{center}
\begin{tikzpicture}[level distance=4em,
  every node/.style = {shape=rectangle, rounded corners,
    draw, align=center}]
    [\tikzstyle{level 1}=[sibling distance=15em],\tikzstyle{level 2}=[sibling distance=8em],\tikzstyle{level 3}=[sibling distance=8em]]]
    \node (a) {non-O, non-null, non-aligned}
    child { node (a0) {choose $\w{k}$ the multiple DP-vector}  
      child { node (b) {$|\kappa|^2+|\sigma|^2\neq 0$}
        child { node (b1) {?}
            edge from parent [->] node [right,draw=none] {}
        }
      }
      child { node (c) {$\kappa=\sigma=0$}
        child { node (c1) {$\Lambda = 0$, $\rho\neq 0$}
          child { node (c11) {$\pi=0$}
            child { node (c112) {Griffiths\cite{Griffiths}}
              child { node (c1121) [label={[draw=none]below: {}}] {$\tau=0\neq \mu$}
                child { node (c11211) [label={[draw=none]below: {}}] {Griffiths\cite{Griffiths1983}}
                edge from parent [->] node [right,draw=none] {}
                }
              }
              child { node (c1122) [label={[draw=none]below: {}}] {$\mu=0\neq \tau$}
                child { node (c11221) [label={[draw=none]below: {}}] {Cahen-Spelkens\cite{CahenSpelkens}}
                  child { node (c1121a) [label={[draw=none]below: {}}] {$\Psi_3 =0$}
                    child { node (c1121b) [label={[draw=none]below: {}}] {Cahen-Leroy\cite{CahenLeroy}}
                         child { node (c1121c) [label={[draw=none]below: {}}] {$\Phi_2\rho-\Phi_0\lambda =0$}
                           child { node (c1121d) [label={[draw=none]below: {}}] {Szekeres\cite{SzekeresJ1966}}
                           edge from parent [->] node [right,draw=none] {}
                           }
                         }
                    edge from parent [->] node [right,draw=none] {}
                    }
                  }
                edge from parent [->] node [right,draw=none] {}
                }
              }         
              child { node (c1123) [label={[draw=none]below: {}}] {\ldots }} 
            edge from parent [->] node [left,draw=none] {\tiny \resfour }
            }
          }
          child [missing] 
          child { node (c12) {$\pi\neq0$}
            child { node (c121) [label={[draw=none]}] {$\Re \rho \neq 0$}
              child { node (c1211) {?}
              }
            edge from parent [->] node [left,draw=none] {\tiny \resfour }  
            }  
          }  
        edge from parent [->] node [left,draw=none] {\tiny \resfour }
        edge from parent [->] node [right,draw=none] {\tiny null-rotate about $\w{k}$ such that $\Phi_1=0$}
        }
      }
    }  
    ;
    \end{tikzpicture}
\end{center} 
}
\caption{\label{fig:nonaligned} {Algebraically special non-nul Einstein-Maxwell solutions for which the multiple Weyl-PND is $\w{k}$ not a PND 
of $\w{F}$. The sub-tree under Griffiths\cite{Griffiths} presents in an invariant way the subcases mentioned in \cite{Griffiths}, 
originally obtained by imposing coordinate restrictions or restrictions on the NP spin-coefficients.}}
\end{figure}
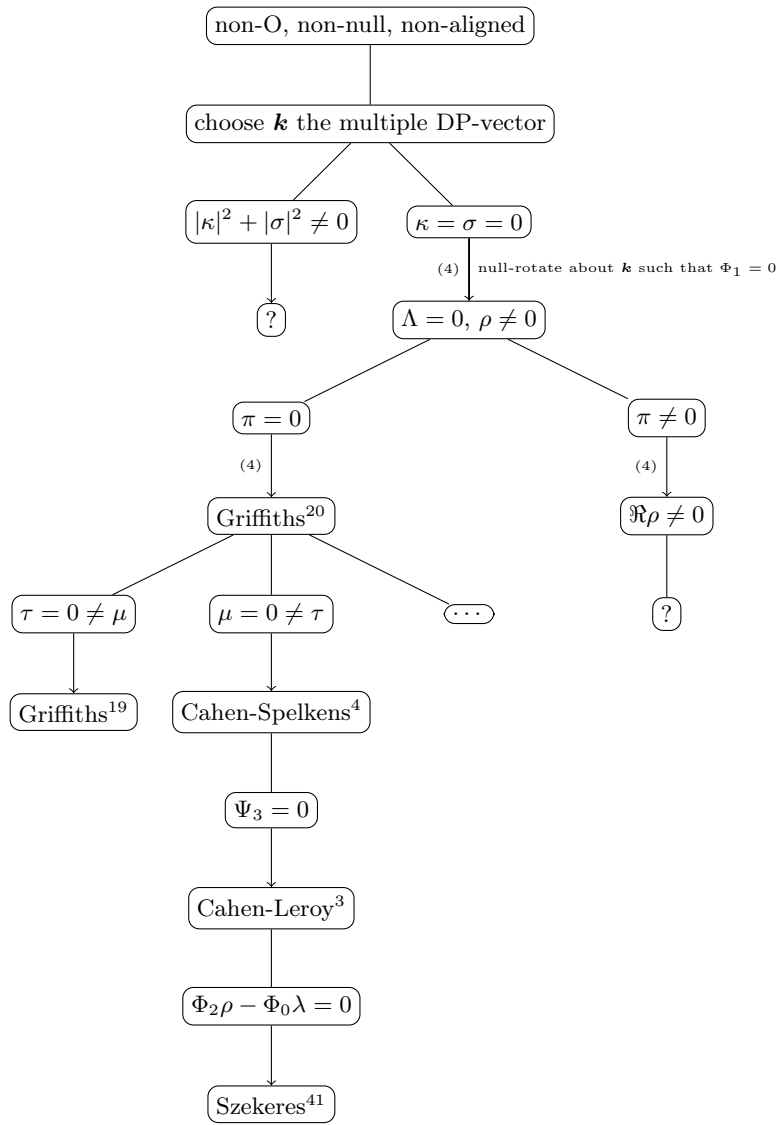

\vfill \eject
\section*{Acknowledgment}
All calculations were done using the Maple symbolic algebra system\ftn{Note3}.
I also thank Lode Wylleman for a critical reading of the manuscript.

\section{Appendix: GHP, Maxwell and Bianchi equations}
Weights
of the spin-coefficients, the Maxwell and Weyl spinor components and the GHP operators:
\begin{eqnarray} 
&\kappa : [3, 1], \nu : [-3, -1], \sigma : [3, -1], \lambda : [-3, 1], \nonumber \\
&\rho : [1, 1], \mu : [-1, -1], \tau : [1, -1], \pi : [-1, 1], \nonumber \\
&\Phi_0 : [2, 0], \Phi_1 : [0, 0], \Phi_2 : [-2, 0], \nonumber \\
&\Psi_0 : [4, 0], \Psi_1 : [2, 0], \Psi_2 : [0, 0], \Psi_3: [-2,0], \Psi_4 : [-4, 0] ,\nonumber \\
&\eth : [1, -1], \etd : [-1,1], \thd : [-1,-1], \tho : [1,1]. \nonumber
\end{eqnarray}

The prime operation is an involution with 
\begin{eqnarray}
 \kappa' &=& -\nu,\sigma'=-\lambda,\rho'=-\mu, \tau'=-\pi,\\
 {\Psi_0}' &=& \Psi_4, {\Psi_1}'=\Psi_3, {\Psi_2}'=\Psi_2,\\
 \Phi_0' &=& -\Phi_2, \Phi_1'=-\Phi_1.
\end{eqnarray}

\noindent The GHP commutators acting on $(p,q)$ weighted quantities are given by:
\begin{eqnarray}
\left[ \tho,\tho' \right] &=& (\pi+\tauc)\et +(\pic+\tau)\etd +(\kappa\nu-\pi\tau +  \frac{R}{24}-\Phi_{11}-\Psi_2)p \nonumber \\ 
&&+(\kc\nuc-\pic\tauc +  \frac{R}{24}-\Phi_{11}-\Pc_2)q,\\
\left[ \et,\etd \right]  &=& (\mu-\muc)\tho  +(\rho-\rhoc)\tho' +(\lambda\s-\mu\rho- \frac{R}{24}-\Phi_{11}+\Psi_2)p \nonumber \\ 
&&-(\overline{\lambda\s}-\muc\rhoc-\frac{R}{24}-\Phi_{11}+\Pc_2)q,\\ 
\left[ \tho,\et \right] &=& \pic\,\tho -\kappa\tho' +\rhoc\,\et +\sigma\etd+(\kappa\mu-\s\pi-\Psi_1)p + (\overline{\kappa\lambda}-\pic\rhoc-\Phi_{01})q .  
\end{eqnarray}

\noindent GHP equations:
\begin{eqnarray}
\tho\rho-\etd\kappa &=& \rho^2+\s\sigmac-\kc\tau+\kappa\pi+\Phi_{00}, \label{ghp1}\\
\tho\s-\et\kappa &=& (\rho+\rhoc)\s+(\pic-\tau)\kappa+\Psi_0, \label{ghp2}\\
\tho\tau-\tho'\kappa &=& (\tau+\pic)\rho+(\tauc+\pi)\s+\Phi_{01}+\Psi_1, \label{ghp3}\\
\tho \nu-\tho' \pi  &=&  (\pi+\tc)\mu+(\pic+\tau)\lambda+\Psi_3+\overline{\Phi_1}\Phi_2,\label{ghp6}\\
\et\rho-\etd\s &=& (\rho-\rhoc)\tau+(\mu-\muc)\kappa+\Phi_{01}-\Psi_1,\label{ghp8}\\
\tho'\s-\et\tau &=& -\s\mu-\lc\rho-\tau^2+\kappa\nuc-\Phi_{02},\label{ghp4d}\\
\tho'\rho-\etd\tau &=& -\muc\rho-\lambda\s-\tau\tauc+\kappa\nu-\frac{R}{12}-\Psi_2. \label{ghp5d}
\end{eqnarray}

\noindent Maxwell equations:
\begin{eqnarray}
\tho \Phi_1-\etd \Phi_0  &=& \pi \Phi_0+2\rho\Phi_1-\kappa \Phi_2, \label{max1}\\
\tho \Phi_2-\etd \Phi_1  &=& -\lambda \Phi_0+2 \pi \Phi_1+\rho\Phi_2. \label{max2}
\end{eqnarray}

\noindent Bianchi equations ($\Phi_{IJ}=\Phi_I\overline{\Phi_J}$ and $\Lambda=R/4=constant$):
\begin{eqnarray}
{\etd} \Psi_{{0}}  -{\tho} \Psi_{{1}}
 +{\tho} \Phi_{{01}}  -{\et} \Phi_{
{00}}  &=& -\pi\,\Psi_{{0}}-4\,\rho\,\Psi_{{1}}+3\,\kappa\,\Psi_{
{2}}+   \pic    \Phi_{{00}}+2\, 
\rhoc    \Phi_{{01}}+2\,\sigma\,\Phi_{{10}} \nonumber \\
&&-2\,\kappa\,\Phi_{{11}}-\kc    \Phi_{{02}}, \label{bi1}\\
{\thd}   \Psi_{{0}}    -{\et}   \Psi_{{1}}
    +{\tho}   \Phi_{{02}}    -{\et}   \Phi_{
{01}} &=& -\mu\,\Psi_{{0}}-4\,\tau\,\Psi_{{1}}+3\,\sigma\,\Psi_{
{2}}-\lc    \Phi_{{00}}+2\,   \pic    \Phi_{{01}}+2\,\sigma\,\Phi_{{11}}\nonumber \\
&&+   \rhoc    \Phi_{{02}}-2\,\kappa\,\Phi_{{12}}, \label{bi2}
\end{eqnarray}
\begin{eqnarray}
&&3\,{\etd}   \Psi_{{1}}    -3\,{\tho}   \Psi_{{2}}
    +2\,{\tho}   \Phi_{{11}}    -2\,{\et}   
\Phi_{{10}}    +{\etd}   \Phi_{{01}}    -{\thd}
   \Phi_{{00}} = 3\,\lambda\,\Psi_{{0}}-9\,\rho\,\Psi_{{2
}}-6\,\pi\,\Psi_{{1}}+6\,\kappa\,\Psi_{{3}}+ (\muc -2\,\mu )   \Phi_{{00}}\nonumber \\
&& \ \  +   2\,(\pi+
   \tauc  )      \Phi_{{01}}+2\,  ( \tau+   \pic  )      \Phi_{{10}}+2\,  ( 2\,
   \rhoc    -\rho )   \Phi_{{11}}
   +2\,\sigma\,\Phi_{{20}
}-\sigmac    \Phi_{{02}}-2\,\kc    \Phi_{{12}}-2\,\kappa\,\Phi_{{21}}, \label{bi3}\\
&&3\,{\thd}   \Psi_{{1}}    -3\,{\et}   \Psi_{{2}}
    +2\,{\tho}   \Phi_{{12}}    -2\,{\et}   
\Phi_{{11}}    +{\etd}   \Phi_{{02}}    -{\thd}
   \Phi_{{01}}  = 3\,\nu\,\Psi_{{0}}-6\,\mu\,\Psi_{{1}}-9
\,\tau\,\Psi_{{2}}+6\,\sigma\,\Psi_{{3}}-\nuc
    \Phi_{{00}}\nonumber \\
&&  \ \ \ +2\, (\muc  -\mu)    
    \Phi_{{01}} -2\,\lc    \Phi_{{10}
}+2\,  ( \tau+2 \pic)        \Phi_{{11
}}+   (2\,\pi+\tauc)        \Phi_{{02}} +   2\,(\rhoc    -\rho)    \Phi_{{12
}}+2\,\sigma\,\Phi_{{21}}-2\,\kappa\,\Phi_{{22}}. \label{bi4}
\end{eqnarray}

\end{document}